\documentclass[conference]{IEEEtran}

\usepackage{amsmath}
\usepackage{cite}
\usepackage{graphicx}
\usepackage{epstopdf} 
\usepackage{multirow}
\usepackage{pbox}
\usepackage{color}
\graphicspath{{./figures/}}
\usepackage{xcolor}
\usepackage[official]{eurosym}
\usepackage[nolist]{acronym} 
\usepackage{amsmath}
\usepackage{algorithm}
\usepackage[noend]{algpseudocode}
\usepackage{comment}

\usepackage{breqn}
\usepackage{amsmath}
\usepackage{adjustbox}
\usepackage{url}

\title{Resource Cooperation in MEC and SDN based Vehicular Networks} 

\author{  
Beiran Chen\IEEEauthorrefmark{1}, Marco Ruffini\IEEEauthorrefmark{1}\\ 
\IEEEauthorblockA{\IEEEauthorrefmark{1} CONNECT centre, School of Computer Science and Statistics, Trinity College Dublin, Ireland, \\\{chenbe, marco.ruffini\}@tcd.ie}
}

\begin{document}

\maketitle


\begin{abstract} 
\ac{IoT} systems require highly scalable infrastructure to adaptively provide services to meet various performance requirements. Combining \ac{SDN} with \ac{MEC} technology brings more flexibility for \ac{IoT} systems. 
We present a four-tier task processing architecture for \ac{MEC} and vehicular networks, which includes processing tasks locally within a vehicle, on neighboring vehicles, on an edge cloud, and on a remote cloud. The flexible network connection is controlled by \ac{SDN}.
We propose a CPU resource allocation algorithm, called \ac{PIRS} with \ac{V2V} communications, based on \ac{ANBS} in Game Theory. \ac{PIRS} encourages vehicles in the same location to cooperate by sharing part of their spare CPU resources.
In our simulations, we adopt four applications running on the vehicles to generate workload. We compare the proposed algorithm with \ac{NCS} and \ac{AIRS}. In \ac{NCS}, the vehicles execute tasks generated by the applications in their own \ac{OBU}, while in \ac{AIRS} vehicles provide all their CPU resources to help other vehicles' offloading requests. 
Our simulation results show that our \ac{PIRS} strategy can execute more tasks on the \ac{V2V} layer and lead to fewer number of task (and their length) to be offloaded to the cloud, reaching up to 28\% improvement compared to \ac{NCS} and up to 10\% improvement compared to \ac{AIRS}.

\end{abstract}


\section{Introduction}
\label{sec:newintroduction}
\ac{MEC} is a technology that extends services to the edge cloud for \ac{IoT} systems.
In MEC-based IoT networks, computational task offloading enhances processing performance for the tasks generated by \acp{UE} of the IoT network. Researchers have been designing offloading strategies to meet diverse performance requirements. 
However, the dynamically changing characteristics of the location and service requests from the \acp{UE} may still lead the fixed edge server deployment to have “service holes” in IoT networks, therefore dynamic communication between the \acp{UE} is necessary
\cite{IoT_MEC_Deep018_17}. 

The vehicle is a type of \ac{UE} in IoT system. Today's vehicles are equipped with \acp{OBU} with multiple sensors, processing units, localization systems, and radio transceivers. These embedded technologies can facilitate the setup of \ac{VANET} \cite{1543745} across vehicles. However, the processing capacity of vehicles is limited, and it is difficult to execute computationally intense tasks within their own \acp{OBU}. Therefore, task offloading to \ac{ES} or cloud is considered as an option to increase the availability of processing power \cite{MEC_tra}. 
SDN-based technologies can be widely adopted in \ac{IoT} system, from different networking aspects, e.g., access, 
edge, core, data center networking \cite{intro_01},
and also useful in \ac{V2V} systems \cite{IoT_MEC_Deep018_19} -\cite{IoT_MEC_Deep018_21}. The SDN controller inside the MEC server can flexibly construct the network topologies between the vehicles, and realize the V2V offloading dynamicity.
Authors in \cite{IoT_MEC_Deep018_19} proposed an architecture using SDN and MEC servers, in which the SDN controller can keep calculating and selecting the best V2V routing path between vehicles.
In \cite{IoT_MEC_Deep018_20}, the authors extend the architecture to multi-hop V2V connection and optimize the path base on the SDN controller deployed in MEC for both V2V and V2I task offloading.
Authors in \cite{IoT_MEC_Deep018_21} proposed a vehicle trajectory prediction model to improve the efficiency of V2V task offloading by utilizing the mobility advantages of vehicles. 
However, these papers didn't consider the willingness of vehicles for resource sharing, since the computational resources are managed in a centralized way as a resource pool.

Our work integrates \ac{V2V}, \ac{V2I}, and \ac{SDN} architecture for task offloading, and extends it with the willingness and cooperation of the vehicles. We propose a Game-Theory-based algorithm to optimize resource allocation. We then investigate four different application types in our simulation, which are typical use cases in the IoT-based vehicular networks with different levels of computational task loads \cite{edgesim_4}.

In this paper, our contributions are as follows:
(1) We propose a four-tier resource cooperation architecture, which uses \ac{SDN} for communication control and \ac{MEC} for task offloading.
(2) We propose a cooperation strategy, dubbed \ac{PIRS}, based on \ac{ANBS} in Game Theory\cite{kalai1977nonsymmetric}, at \ac{V2V} layer to reallocate the spare resource of each vehicle, which considers vehicles' cooperation history and willingness.
(3) We simulate the task offloading schemes (with cutting-edge simulator EdgeCloudSim \cite{edgesim_1}) and benchmark our \ac{PIRS} strategy against two other strategies: non-cooperative \ac{NCS} strategy and cooperation with all idle resource (\ac{AIRS}) strategy. The results show an obvious performance advantage of our strategy. 


\section{System Model}
\label{sec:systemModel}

\subsection{Architecture}
\label{sec:Architecture description}
\begin{figure}[h]
    \vspace{-0.1in}
    \centering
\includegraphics[width=0.45\textwidth]{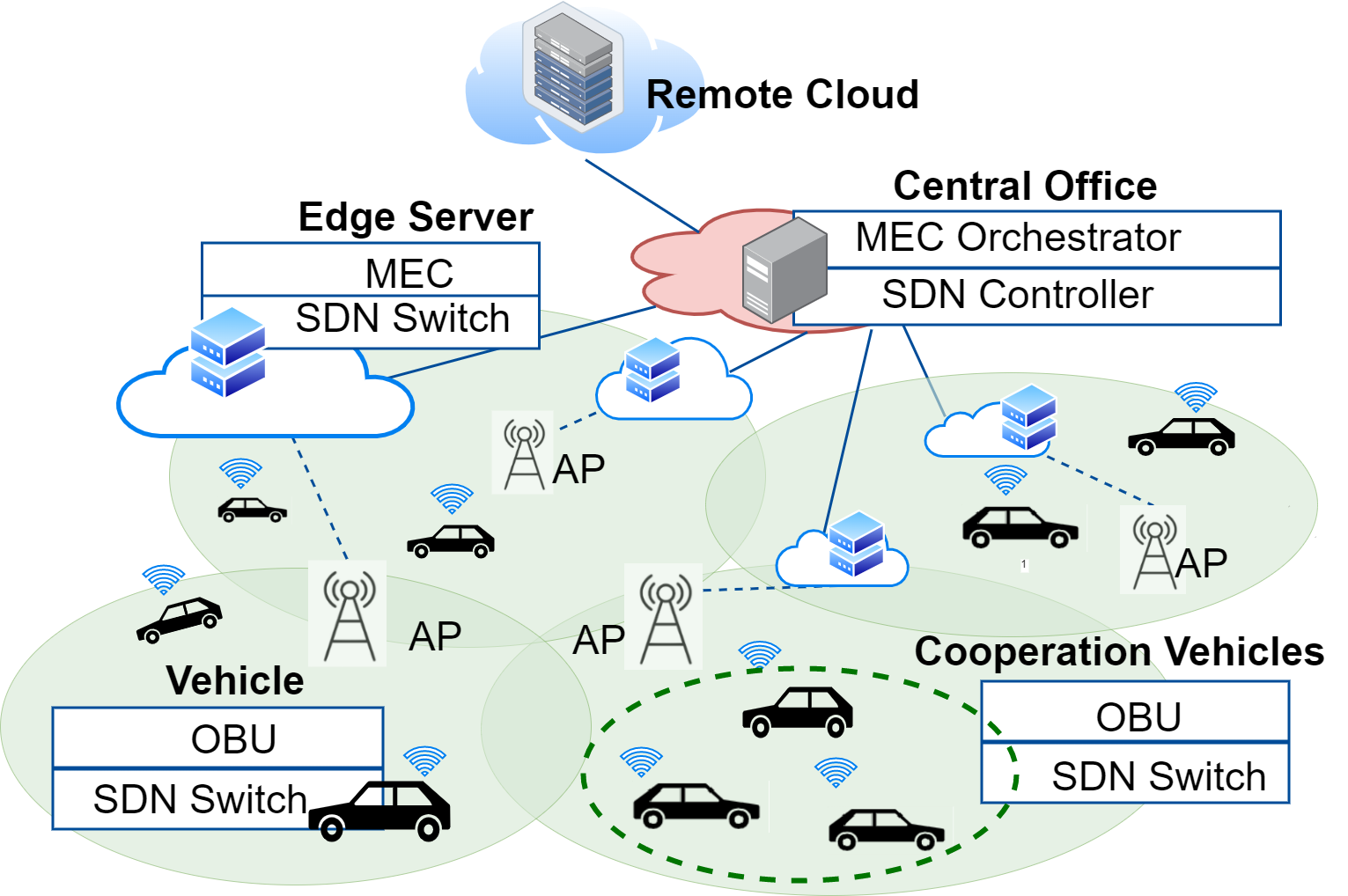}
    \caption{Proposed MEC architecture.}
    \label{fig:whole_architecture}
    \vspace{-0.1in}
\end{figure}

Our proposed architecture is shown in Fig. \ref{fig:whole_architecture}. It's a \ac{SDN}-based four-tier architecture, which includes processing at local vehicle on-board, neighboring vehicles, edge cloud, and remote cloud. 
We assume each vehicle is equipped with \ac{OBU}, and has a certain computational ability. The tasks generated by vehicles are prioritized to be executed in the local on-board CPU (i.e., the \ac{OBU}) first. If the on-board CPU capacity is not sufficient, it cooperates/negotiates with neighboring vehicles by \ac{V2V} communications, gets resources from them, and sends the remaining task to them to execute. After that, if still not sufficient,
the remaining task is offloaded to \acp{ES} by \ac{V2I} communication. Finally, the last option is to offload the remaining task to the remote cloud, in case \acp{ES} get congested and do not have enough computational resources, especially when there are many demands coming from a large number of vehicles for the \ac{ES}. 
Each \ac{ES} is associated with an \ac{AP}.
In our architecture, V2V connections use the IEEE 802.11p standard, while V2I connections use the IEEE 802.11ac standard \cite{edgesim_4}. These connections are controlled by \ac{SDN}. Every vehicle, and \ac{ES} has an \ac{SDN} switch. 
All the connection establishment between them and data transmission are controlled by the \ac{SDN} controller located in the central office. 
The procedure of task offloading is handled by the \ac{MEC} orchestrator, located in the central office as well, adopting architectures such as those defined in \cite{Das_JOCN}. 

For modeling the mobility of vehicles, we divide the whole map into several areas by \ac{AP} coverage. When vehicles drive in an \ac{AP} coverage area for a short period and move out of this area to another location, we define this short period as a dwell time. Different locations are assumed to have different levels of dwell time for the vehicles since different areas have different average driving speeds.
We have a randomly distributed vehicle generator to map vehicles into \acp{AP}, and a dwell time to simulate the mobility of vehicles in areas covered by different \acp{AP}. Vehicles are assumed to move out of their \ac{AP} coverage area after the dwell time has expired and move into an adjacent \ac{AP} coverage area for a new dwell time.

\vspace{-0.1in}

\subsection{Offloading strategy}
As mentioned above, in our architecture, when a task is generated it has four ways to be executed: locally in the vehicle, across neighboring vehicles, on the \ac{ES}, and on the remote cloud. 

\subsubsection{Local task execution}
The strategy where a vehicle $V_{i}$ always executes the task $K(L_{K},D_{K})$ on its own \ac{OBU} without cooperation with other vehicles, is named \ac{NCS}. The task length $L_{K}$ denoted by the number of instructions, $D_{K}=(D^{up}_{K},D^{down}_{K})$ is the task upload/download data size. 
We assume every vehicle has the same CPU capacity and the spare CPU resource of the vehicle $V_{i}$ at time $t$ is represented by $C_{i}(t)$. 
The execution time $d_{c}^{l}(K)$ is determined by its current spare CPU capacity  $C_{i}(t)$. Since there is no communication delay, the delay $d^{l}(K)$ only includes the computational delay, shown in Equation (\ref{eqn:delay_noncooperation}). 
The unfinished tasks are directly offloaded to \ac{MEC}.

\vspace{-0.05in}
\begin{equation}
    \vspace{-0.05in}
d^{l}(K) = d_{c}^{l}(K) = L_{K} / C_{i}(t)
\label{eqn:delay_noncooperation}
\end{equation}

\subsubsection{Cooperative task execution with neighboring vehicles}
When a vehicle $V_{i}$ finds that the local estimated delay $d^{l}(K)$ is larger than the task's delay tolerance $d_{limit, K}$, the vehicle carries out a new estimation of the \ac{V2V} delay $d^{g}(K)$ if the processing were to be executed on a cooperating neighbouring vehicle. In our scheme, the cooperation has the following steps:

\textbf{Geographical grouping}: Vehicles are grouped by their geographical locations. The \ac{SDN} controller at the central office collects the information of the vehicle $V_{i}$'s neighbors' geographical region and spare CPU resource $C_{j}(t)$, and sends the neighbor set  $N = \left \{ N_{1}, N_{2} ...,N_{j}\right \}$ to the task owner vehicle $V_{i}$. 
All this information is useful for selecting cooperating neighboring vehicles.

\textbf{Utility Equation}: Each vehicle $i$ evaluates its utility value when it cooperates with other vehicles. Our utility equation is defined in  Equation (\ref{eqn:v_i})\cite{br-swarm}. 
It considers the current environment state and the vehicles' willingness to cooperate. The current environment state of vehicle $V_{i}$ ($i$ is the ID index of vehicles) at time $t$, is denoted by $s=[\theta_{0},\theta_{1}]$. $\theta_{0}$ means the vehicle $V_{i}$ has spare resources, while $\theta_{1}$ means it does not. 
The risk probability vector $P\theta_{i}=[P\theta_{i}(\theta_{0},t), P\theta_{i}(\theta_{1},t)]$ represents the probability of the vehicle in a risky (the vehicle might not have enough computing resources left for its own tasks) or safe state (the vechicle does not risk to run out of resources). $ P\theta_{i}(\theta_{0},t)$ denotes the probability of the vehicle $V_{i}$ in risky state $\theta _{0}$, and $P\theta_{i}(\theta_{1},t)$ denotes the probability of the vehicle in safe state $\theta _{1}$, defined by Equations (\ref{eqn:P_risk}). 
This equation includes two terms: the first term is the real-time usage $B_{i}(t)$ spent on real-time resource $C^{r}_{i}(t)$. The real-time resource $C^{r}_{i}(t)$ is the total resource vehicle $V_{i}$ has at time $t$. The real-time spare CPU resources are defined as $C_{i}(t) = C^{r}_{i}(t) - B_{i}(t)$.

The second term represents the willingness of vehicle $V_{i}$ to join the current round of cooperation. The cooperation willingness probability vector is $\beta_{i}(t)=[\beta _{i,0}(t),\beta _{i,1}(t)]$, $(\beta _{i,0}(t)\geq 0,\beta _{i,1}(t)\geq 0)$. Here $\beta _{i,0}(t)$ is the `giving' probability, which denotes the vehicle's willingness to give its resource to other vehicles at a given time, while $\beta _{i,1}(t)$ is the `getting' probability, which denotes the vehicle's intention to get resources from other vehicles within that same time. Note that $\beta _{i,0}(t)+\beta _{i,1}(t)=1$ \cite{Givigi-swarm}, i.e., vehicles are not allowed to both give and get resources within the same time window.
The willingness probability vector $\beta_{i}(t)$ changes after each cooperation round, and depends on all previous cooperation rounds of $V_{i}$. 


\vspace{-0.05in}
\begin{equation}
\vspace{-0.1in}
\label{eqn:v_i}
\begin{split}
U_{i}\left ( s,a,t \right ) = E\left \{J_{i}\left ( s ,a ,t \right )  \right \}   
=M\theta \cdot P\theta_{i}^T \cdot J_{i}(s,a ,t)\\= \begin{bmatrix}
    M\theta_{11}, M\theta_{12}\\ 
    M\theta_{21}, M\theta_{22} 
\end{bmatrix}\cdot \begin{bmatrix}
    P\theta_{i}(\theta_{0},t)\\ 
    P\theta_{i}(\theta_{1},t)
\end{bmatrix}\cdot J_{i}(s,a,t) \\
\quad \left (i,j=1,2....n\quad i\neq j \right )\\
\end{split}
\end{equation}
Where, $P\theta_{i}$ and $J_{i}(s,a,t)$ are derived by the following Equations\cite{br-swarm}:
\vspace{-0.2in}

\begin{equation}
\vspace{-0.1in}
\begin{split}
P\theta_{i}(\theta_{0},t)=\frac{B_{i}(t)}{C^{r}_{i}(t)} +\left ( \beta _{i,0}(t)-\beta _{i,1}(t) \right ) ;\quad\quad\quad\quad\quad\quad\quad\\
   P\theta_{i}(\theta_{1},t)=1-P\theta_{i}(\theta_{0},t) \quad\quad\quad\quad\quad\quad\quad
    \label{eqn:P_risk}
    \end{split}
\end{equation}

\begin{equation}
\label{eqn:J_i}
\begin{split}
J_{i}(s,a,t)  =Pa_{i}\cdot Ma\cdot Pa_{j}^T \quad\quad\quad\quad\quad\quad\quad\quad\quad\quad\quad\quad\\
 =[Pa_{i}(a_{0},t), Pa_{i}(a_{1},t)]\cdot \begin{bmatrix}
    Ma_{11},Ma_{12}\\ 
    Ma_{21},Ma_{22} 
    \end{bmatrix}\cdot \begin{bmatrix}
    Pa_{j}(a_{0},t)\\ 
    Pa_{j}(a_{1},t)
\end{bmatrix} \\
\end{split}
\end{equation}
The factor $J_{i}(s,a,t)$ in Equation (\ref{eqn:J_i}) reflects the reward that the vehicle can get from its current action \cite{br-swarm}. We define a vehicle $V_{i}$ action space as $a =[a_{0},a_{1}]$ having two action choices, giving out resources to help others, denoted as $a_{0}$, or getting resources from others, denoted as $a_{1}$. The giving probability vector $Pa_{i}=[Pa_{i}(a_{0},t), Pa_{i}(a_{1},t)]$ represents the probability of the vehicle selecting the give/get action. $Pa_{i}(a_{0},t)$ is the probability that the vehicle gives out its resource. $Pa_{i}(a_{1},t)$ is the probability that the vehicle chooses to get other's resource. In our case, the task owner vehicle joining the cooperation selects $a_{1}$, and its giving probability $Pa_{i} = [0,1]$, while its neighboring vehicle who takes part in the cooperation selects $a_{0}$, and $Pa_{i}= [1,0]$. 
Based on Game Theory, the payoff matrix $Ma$ and $M\theta$ are defined empirically. The vehicles are encouraged to be rewarded for cooperating with each other to execute the tasks. Therefore, the Nash Equilibrium point, in this case, will be reached when the neighboring vehicles prefer to form a coalition with the task owner vehicle without getting into a risk environment which might lead to a lack of CPU resources to process their own tasks.
The central office gives a list of neighboring vehicles sorted by their utilization values to the task owner vehicle $V_{i}$.

\textbf{Selecting and cooperating with neighboring vehicles:} After the task owner vehicle $V_{i}$ gets the cooperation list $N^c$ from the central office, it selects the top utility value neighbors to be cooperating candidates to execute the task $K(L_{K},D_{K})$, which is generated by vehicle $V_{i}$ at the time $t$.
The $V_{i}$ estimates the delay time $d^{g}(K)$ of the offloading task to those cooperating candidates. This is represented in Equation (\ref{eqn:v2v_delay}). The delay $d^{g}(K)$ includes communication delay $d^{g}_{m}(K)$ between vehicles and computational delay $d^{g}_{c}(K)$.
We assume the tasks in the \ac{V2V} layer can be partitioned. We denoted the communication data rate with $b_{V}$. The vehicle $V_{i}$ selects $min (N^c, N^n)$ neighbors to cooperate. $N^n$ is the maximum number of vehicles that one vehicle can connect to. 

If the estimated execution time $d^{g}(K)$ is less than the task's maximum tolerable time $d_{limit, K}$, the task $K(L_{K},D_{K})$ will be executed on the \ac{V2V} layer. 
In our scheme, each vehicle is an \ac{SDN} switch and is controlled by the \ac{SDN} controller located in the central office. 
The \ac{SDN} controller can build a temporary connection when the vehicle $V_{i}$ forms a coalition with selected cooperation candidates. This coalition is temporary. It is formed when the vehicles cooperate to execute a task, and it's cancelled when the task is finished. 
In Equation (\ref{eqn:v2v_delay}), $L_{K}$ denotes the task length and $D^{up}_{K}$/$D^{down}_{K}$ represents task upload/download data size. $C_{i}(t)$ is the spare CPU resource of $V_{i}$, while $\sum_{j=1}^{min(N^c, N^n)} C_{j}(t)$ is the sum of resources provided by each cooperating neighboring vehicle. 
  	\vspace{-0.05in}
\begin{equation}
   	\vspace{-0.05in}
\label{eqn:v2v_delay}
\left\{\begin{matrix}
d^{g}(K)= d^{g}_{c}(K) + d^{g}_{m}(K)\quad\quad\quad\\
d^{g}_{c}(K)= \frac{L_{K}}{ C^{r}_{i}(t)}\quad\quad\quad\quad\quad\quad\quad\quad\\
C^{r}_{i}(t)= C_{i}(t) + \sum_{j=1}^{min(N^c, N^n)} C_{j}(t) \\
d^{g}_{m}(K)=\frac{D^{up}_{K}}{b_{V}} +  \frac{D^{down}_{K}}{b_{V}}\quad\quad\quad\quad\\
\end{matrix}\right.
\end{equation}

We assume a cooperating neighboring vehicle $N_{j}$ has $C_{j}(t)$ spare resource value at the time $t$. If the vehicle $N_{j}$ provides all of its spare resources to help process $V_{i}$'s offloading task at current time $t$, we call this reallocation algorithm \ac{AIRS}. However, the drawback of the \ac{AIRS} approach is that vehicle $N_{j}$ might become unable to process its own upcoming tasks so that it has to offload them to other vehicles or even to \ac{ES}/remote cloud, which would have cost implications.

Here we propose a \ac{PIRS} algorithm, based on \ac{ANBS} in Game Theory\cite{kalai1977nonsymmetric}, where neighboring vehicle $N_{j}$ provides part of its spare resources for cooperation. The task owner vehicle $V_{i}$ cooperates with neighboring vehicles $N_{j}$ in the candidate list $N^{c}$ one by one in descending order of their utility values.
 In each cooperation round, candidate neighboring vehicle $N_{j}$, which adopts the \ac{PIRS} algorithm does not provide all its spare resources for cooperation, but only part of it to process task owner vehicle $V_{i}$'s offloading.
We consider this cooperation as a bargain problem and assume both vehicles are rational and intend to maximize their extra resource utility in the bargain.

The set of spare resources in the utility equation in this bargain problem can be described as $\Gamma=\left \{ \gamma_{i}\left |i=1,2  \right \} \right \}$, where $\gamma _{i}=\left \{ (C^{r}_{i}(t)-B_{i}(t))|i=1,2  \right \}$, which is a nonempty compact convex set with boundary \cite{MatrainANBS}. $C^{r}_{i}(t)$ is the real-time total CPU resource for each vehicle, including its own CPU resource and the resource it gets externally, while $B_{i}(t)$ is the real-time resource usage.
The cooperation problem is described in Equation (\ref{eqn:ANBS_two}):
\begin{equation}
\vspace{-0.1in}
    \label{eqn:ANBS_two}
    \begin{split}
\Gamma ^{*}=\underset{C^{r}_{i}(t)}{argmax}\prod_{i}(C^{r}_{i}(t)-B_{i}(t))^{\lambda _{i}(t)},\\
s.t. \quad\quad\quad\sum_{i=1}^{2}C^{r}_{i}(t)=\Phi \\
 \sum_{i=1}^{2} \lambda _{i}(t)=1\\
 C^{r}_{i}(t)\geq B_{i}(t),\\
 C^{r}_{i}(t)\geq 0, \quad (i=1,2) 
    \end{split}
\end{equation}
\vspace{+0.2in}

where $\Phi$ is the total real-time resource of the vehicles considered. 
\begin{equation}
    \label{eqn:lambda}
\lambda _{i}(t)=\frac{\beta _{i,1}(t)}{\beta _{i,1}(t)+\beta _{j,1}(t)}\quad\left ( i,j=1,2,3...n \right )
\end{equation}
$\lambda _{i}(t)$ denotes the bargaining power of the vehicles.
In our case, the vehicles' bargaining power\cite{MatrainANBS} is decided by their willingness probability vector $\beta_{i}(t)$ as Equation (\ref{eqn:lambda}).
At each allocation step, the vehicle gets its available real-time resource as $C^{r}_{i}(t)=B_{i}(t)+\lambda_{i}(t) \cdot \sum_{i=1}^{2}{(C^{r}_{i}(t)-B_{i}(t))}$.

After the cooperation, the algorithm updates the parameters of the cooperating vehicles. Part of the spare resources of neighboring vehicle $N_{j}$ are provided to execute $K(L_{K},D_{K})$ offloaded from $V_{i}$,
thus $V_{j}$'s risk probability of lacking CPU resources increases. 
Therefore, the cooperation willingness probability vector $\beta_{i}(t)$ changes. 
In addition, the cooperation also changes the participants' bargaining power $\lambda_{i}(t)$, which will affect their next round of cooperation. We use the updating rule derived from \cite{Multiagentbook}, as the Equation (\ref{eqn:update_A_trait}) - Equation (\ref{eqn:trait_A_delta}).

\begin{multline}
\vspace{-0.2in}
    \label{eqn:update_A_trait}
\quad \quad \beta_{i,m}(t) =\beta_{i,m}(t-1) +\alpha \Delta \beta_{i,m}(t),\\
\quad \left(m=0,1\right ),\quad  \alpha \in (0,1) 
\end{multline}

where, $\alpha$ is the learning rate and the $\Delta \beta_{i,m}(t)$ holds as:

\begin{multline}
\vspace{-0.1in}
\label{eqn:trait_A_delta}
 \quad\quad  \Delta \beta_{i,m}(t)=\frac{\Delta J_{i}(s,a_{m},t)}{\Delta J _{i}(s,a_{m},t)+\Delta J _{i}(s,a_{l},t)},\\
\Delta J _{i}(s,a_{m},t)=J _{i}(s,a_{m},t)-J _{i}(s,a_{m},t-1),
\\ \quad \left (m,l=0,1\quad m\neq l  \right )
\end{multline}


If the estimated execution time $d^{g}(K)$ is more than the task's maximum tolerable delay $d_{limit, K}$, the task $K(L_{K},D_{K})$ will be offloaded to \ac{ES}/remote cloud.

\subsubsection{Offloading to Edge or remote cloud}
When vehicles decide to offload tasks to \ac{ES}/remote cloud, they communicate with their nearest \ac{ES}. 
When the \ac{ES} is congested, tasks can be offloaded to the remote cloud. 
The delay $d^{e}(K)$ of \ac{V2I} includes communication and computational delay which is determined by the computational capacity of offloading \ac{ES}, shown in Equation (\ref{eqn:V2I_delay}).

\begin{equation}
\label{eqn:V2I_delay}
\left\{\begin{matrix}
d^{e}(K)= d^{e}_{c}(K) + d^{e}_{m}(K)\\
d^{e}_{c}(K)= \frac{L_{K}}{ C_{E}} \quad\quad\quad\quad\quad\\
d^{e}_{m}(K)=\frac{D^{up}_{K}}{b_{E}} +  \frac{D^{down}_{K}}{b_{E}}\quad
\end{matrix}\right.
\end{equation}
where, $L_{K}$ denotes the task length and $D^{up}_{K}$/$D^{down}_{K}$ represents task upload/download data size. $C_{E}$ is the CPU resource provided by \ac{ES}/remote cloud.

In summary, the whole procedure of our proposed algorithm \ac{PIRS}, as well as \ac{AIRS} and \ac{NCS}, are shown in Algorithm \ref{alg:v2v}.
The computational complexity of algorithm \ac{PIRS} and \ac{AIRS} is mostly affected by the sorting algorithm in Line 13, in which the algorithm gets candidates list $N^c$ by sorting candidates' utility value Equation (\ref{eqn:v_i}). Our code adopts Python built-in Timesort algorithm \cite{python_sort}, and the complexity is $O(nlog n)$. The computational complexity of \ac{NCS} is $O(1)$ since it does not have sharing and offload all remaining tasks to MEC.
\begin{algorithm}[h]
\caption{Four-tier offloading system with \ac{V2V} cooperation algorithm \ac{PIRS}, \ac{AIRS} and \ac{NCS}}
\label{alg:v2v}
\begin{algorithmic}[1]
\State Initialization: task $K(L_{K},D_{K})$ generated by vehicle $V_{i}$
\If{strategy is \ac{NCS}}
\State estimate delay $d^{l}(K)$ Eqn.\ref{eqn:delay_noncooperation}
    \If{$ d^{l}(K)  < d_{limit, K}$} 
        \State executes the task locally
    \Else
        \State the task failed
    \EndIf
\EndIf
\If{strategy is \ac{PIRS} or \ac{AIRS}}
    \State Offloading strategy:
    \State \ac{V2V} cooperate to execute the task
    \State1) get geographical neighbor set $N$ 
    \State2) select candidates list $N^c$ from neighbor set $N$,by sorting their utility value Eqn.\ref{eqn:v_i}
    \State3) calculate the total resource  $V_{i}$ can get from neighbors:

    \For{neighbor $N_{j}$ in candidate list $N^c$}
        \If {Strategy is \ac{PIRS}}
         \State $N_{j}$ gives a part of its spare resource Eqn.\ref{eqn:ANBS_two}
         \State update willingness probability $\beta_{i}(t)$, Eqn.\ref{eqn:update_A_trait}-\ref{eqn:trait_A_delta}
        \ElsIf {Strategy is \ac{AIRS}}
         \State $N_{j}$ gives all spare resource
        \EndIf
  \EndFor
    4) calculate the delay $d^{g}(K)$, Eqn.\ref{eqn:v2v_delay}
    \If{$d^{g}(K) < d_{limit, K}$}
    \State V2V cooperates to execute the task
    \Else
    \State Offloading to Edge/remote cloud
        \If {\ac{VM} utilization $<$ utilization threshold} 
            \State offload to nearest \ac{ES}
        \Else
            \State offload to remote cloud
        \EndIf
    \State  calculate the delay $d^{e}(K)$, Eqn.\ref{eqn:V2I_delay}
    \EndIf
    
  \If{$d^{e}(K) < d_{limit, K}$ OR $V_{i}$ change place } 
        \State the task failed
  \Else    
        \State the task successfully executed
  \EndIf 
  
\EndIf 

\end{algorithmic}
\end{algorithm}

\section{Experimental results}
\label{sec:results}
\subsection{Simulation settings}
We implement our own Python-based simulator for the \ac{V2V} part and use an open-source Java-based simulator, EdgeCloudSim \cite{edgesim_1}, for the \ac{V2I} part, and then integrate them together.
In our simulations, the task execution can fail for two reasons. The first reason is the mobility of vehicles. 
If the vehicle moves out of the wireless network coverage, it is not connected to the previous \ac{ES} anymore, and it cannot get the response of its previously requested task.
The second reason is the delay. If a task execution cannot finish within its maximum tolerable delay, it fails.
In our simulation, we adopt the parameters of four task applications in paper \cite{edgesim_4} as our use cases, shown in Table \ref{tab:v2v_taskParameter}. 
The usage percentage of the application is defined as the proportion of the vehicles running this application. The task inter-arrival time means how frequently a given task generates a processing load. This inter-arrival time is exponentially distributed\cite{edgesim_4}. 
The maximum tolerable delay is the time limit for the task finishing time. If the task execution time goes beyond it, the task fails.
There are also active/idle periods for generating the task. During the active period, applications generate tasks with the aforementioned inter-arrival time, while during the idle period, applications do not generate any processing load.
The upload/download data size is the communication data size for the task when it is offloaded to other vehicles or to the edge/remote cloud.
The task length represents the task computational quantity and is also an exponentially distributed random variable\cite{edgesim_4}.
The \ac{VM} utilization denotes the CPU overhead on the \ac{VM} when it is running on \ac{ES}.
Other parameters of the configuration are listed in Table \ref{tab:edgeParameter}. We adopt EdgeCloudSim's built-in nomadic mobility model for our vehicles. In this model, different locations have different values to represent the different average dwell times the vehicles spend at these locations. In our simulation, we set three types of locations with different average dwell times. We use the EdgeCloudSim default value to set the \ac{ES} layer and remote cloud computational capacity, as well as network communication data rates.

\begin{table}[]
 \caption{Application parameters}
 \vspace{-0.1in}
 \label{tab:v2v_taskParameter}
\begin{adjustbox}{width=\columnwidth, center}
\centering
\begin{tabular}{|l|llll|}
\hline
                                                                         & \multicolumn{1}{l|}{Augmented Reality}   & \multicolumn{1}{l|}{Health App}   &\multicolumn{1}{l|}{Compute Intensive}    & \multicolumn{1}{l|}{Infotainment App}    \\ \hline
Usage percentage(\%)                                                     &\multicolumn{1}{l|}{ 30}    & \multicolumn{1}{l|}{20}      &\multicolumn{1}{l|} {20}       &\multicolumn{1}{l|} {30}       \\ \hline
Task arrival poison  mean (s)                                                  &\multicolumn{1}{l|}{ 1}       &\multicolumn{1}{l|}{ 1 }      &\multicolumn{1}{l|}{10 }      &\multicolumn{1}{l|}{ 5}        \\ \hline
Maximum tolerable delay (s)                                                    &\multicolumn{1}{l|}{ 5 }      & \multicolumn{1}{l|}{8}       & \multicolumn{1}{l|}{8}        &\multicolumn{1}{l|}{ 1}        \\ \hline
Active/Idle Period (s)                                                 &\multicolumn{1}{l|}{40/5}    &\multicolumn{1}{l|} {45/90}   & \multicolumn{1}{l|}{60/120}   & \multicolumn{1}{l|}{30/45}    \\ \hline
\begin{tabular}[c]{@{}l@{}}Upload/Download \\ Data size(KB)\end{tabular} &\multicolumn{1}{l|}{ 1500/25} & \multicolumn{1}{l|}{1250/20} & \multicolumn{1}{l|}{2500/200} & \multicolumn{1}{l|}{2500/200} \\ \hline
Task Length (GI)                                                         &\multicolumn{1}{l|}{9}       & \multicolumn{1}{l|}{3}       & \multicolumn{1}{l|}{45}       &\multicolumn{1}{l|}{45}       \\ \hline
\ac{VM} Utilization on Edge (\%)                                              & \multicolumn{1}{l|}{6}       &\multicolumn{1}{l|}{ 2}       & \multicolumn{1}{l|}{30}       & \multicolumn{1}{l|}{30}       \\ \hline
\end{tabular}
\end{adjustbox}
\vspace{-0.1in}
\end{table}


\begin{table}[]
 \caption{Simulation parameters}
 \vspace{-0.1in}
 \label{tab:edgeParameter}
\begin{adjustbox}{width=0.85\columnwidth, center}
\centering
\begin{tabular}{|l|l|}
\hline
Parameter                                                                        & Value            \\ \hline
Simulation Time                                                                  & 30 minutes       \\ \hline
WAN data rate                                                             & 1 Gbps        \\ \hline
V2I communication data rate                                                             & 250 Mbps     \\ \hline
V2V communication data rate                                                & 10 Mbps \\ \hline
CPU capacity per  Vehicles/Edge/Remote Cloud                                                      & 2/160/1600 GIPS      \\ \hline
Maximum number of V2V connection $N^n$                                                  & 6          \\ \hline
Number of locations Type 1/2/3                                                   & 1/1/2            \\ \hline
Average dwell time in Type 1/2/3                                                 & 30/20/10 seconds \\ \hline

\end{tabular}
\end{adjustbox}
\vspace{-0.2in}
\end{table}


\subsection{Simulation results}

\begin{figure}[h]
\vspace{-0.1in}
    \centering
    	\includegraphics[width=0.39\textwidth]{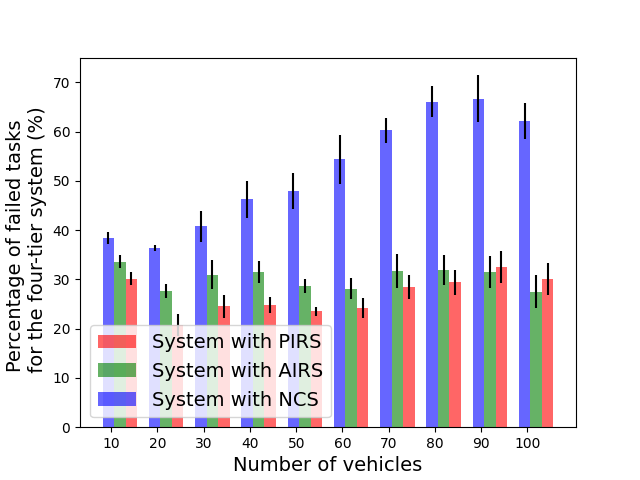}
    \caption{Failed tasks normalized percentage with the four-tier system. }
    \label{fig:V2V+Edgeunfinished}
      	\vspace{-0.1in}
\end{figure}

We investigate the performance of our \ac{PIRS} algorithm and compare it to two baselines \ac{V2V} algorithms: \ac{AIRS} and \ac{NCS}. Our results, which include mean and standard deviation, are shown in the following 4 plots.
Firstly, we investigate the amount of failed tasks for those three systems. As mentioned before the reasons for failed tasks are the mobility of vehicles and exceeding the tolerable delay. Fig. \ref{fig:V2V+Edgeunfinished} shows the normalized failed task percentage for the three systems. The systems adopting \ac{PIRS} and \ac{AIRS} have lower failed task percentages than the system using \ac{NCS}. The system with \ac{PIRS} has the best performance. 
When the vehicle number is higher, the advantage of \ac{PIRS} over \ac{AIRS} is reduced, but the advantage of \ac{PIRS} over \ac{NCS} increases. Please note that, in order to better show the performance comparison, we use aggressive parameter settings (i.e., very frequent task inter-arrival time) in order to increase the overall probabilities of failed tasks. 

\vspace{-0.10in}
\begin{figure}[h]
\vspace{-0.15in}
    \centering
    	\includegraphics[width=0.39\textwidth]{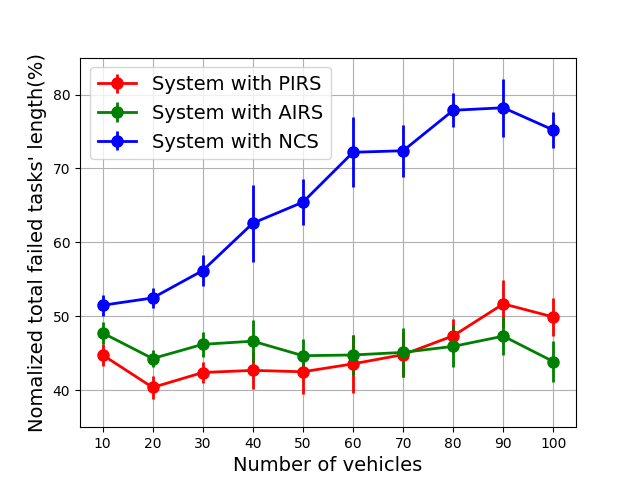}
    	  	\vspace{-0.1in}
    \caption{Normalized total failed tasks' task length. }
    \label{fig:system failed task length}
    	\vspace{-0.1in}
\end{figure}
\begin{figure}[h]
      	\vspace{-0.1in}
    \centering
    	\includegraphics[width=0.39\textwidth]{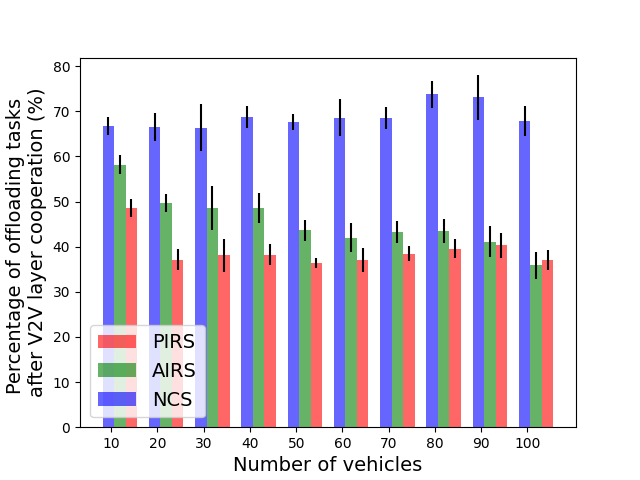}
    \caption{ Offloading task percentage.}
    \label{fig:unfinished}
      	\vspace{-0.1in}
\end{figure}
In Fig. \ref{fig:system failed task length}, we analyze the total length of failed tasks, and we can see the system with the \ac{PIRS} leaves the smallest amount of computational task length uncompleted when the vehicle number is less than 80. In other words, this system executed the highest amount of computations successfully. The system with the \ac{AIRS} has a more uncompleted computational task length than \ac{PIRS} but less than the system with the \ac{NCS}. The advantage of \ac{PIRS} is more obvious with lower density of vehicles. The reason is that the \ac{AIRS} algorithm makes the neighbor vehicles $N_{j}$ provide all of their spare CPU resources at each cooperation, while \ac{PIRS} takes only part of $N_{j}$ spare resources. The proposed \ac{PIRS} algorithm thus provides a more fair distribution in the usage of vehicles' computational resources. This is especially useful for a low number of vehicles, because if a vehicle provides all its computational capacity to another vehicle, it would then have to offload its own task to other vehicles, but there might not be any vehicle nearby. When the number of vehicles becomes higher, there are more options for offloading to other vehicles, thus the performance of \ac{PIRS} and \ac{AIRS} show less difference. 


      	\vspace{-0.05in}
\begin{figure}[h]
\vspace{-0.1in}
    \centering
    	\includegraphics[width=0.39\textwidth]{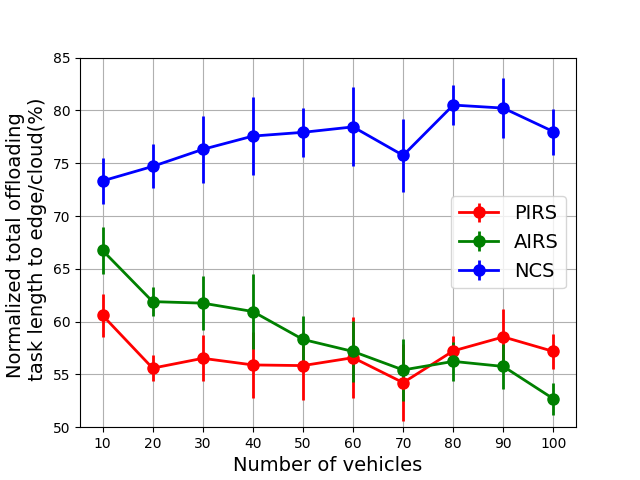}
    	\vspace{-0.1in}
    \caption{Normalized total offloading task length. }
    \vspace{-0.15in}
    \label{fig:offloading task length}

\end{figure}
In order to understand how much the \ac{V2V} resource sharing helps for the whole system performance, we also investigate the first \ac{V2V} layer performance separately.
Fig. \ref{fig:unfinished} shows the percentage of remaining tasks, which need to be offloaded to the MEC layer and remote cloud after the three different algorithms reallocate CPU resources in the \ac{V2V} layer. Compared to the \ac{NCS} and \ac{AIRS} algorithms, \ac{PIRS} can complete more tasks in the \ac{V2V} layer, thus offloading fewer tasks to the MEC layer and cloud. The advantage is around 20\% to 30\% for different scenarios of vehicle densities.
Fig. \ref{fig:offloading task length} shows the total task length offloaded to \ac{ES}/remote cloud (this includes the potential failed tasks). We can see that \ac{PIRS} has the shortest task lengths offloaded to \ac{ES}/remote cloud when the number of vehicles is less than 80. In the best case, the average number of tasks offloaded by \ac{PIRS} with 40 vehicles is 10\% lower than \ac{AIRS}, and 28\% lower than \ac{NCS}, respectively. The reason is the same as mentioned above.
When the number of vehicles increases over 80, even though a vehicle gives out all of its spare resources to $V_{i}$ at the previous cooperation, it has a better chance to group with another vehicle, which has adequate spare resources for its future upcoming tasks. Therefore in this scenario, \ac{PIRS} does not have much advantage compared to \ac{AIRS}.
Finally, both \ac{PIRS} and \ac{AIRS} perform better than the non-cooperation case \ac{NCS} where there is no resource sharing between vehicles.


\section{Conclusion}
\label{sec:newconclusion}
We have proposed a four-tier architecture for vehicular networks, with flexible network connection controlled by \ac{SDN} in V2V communication. We implement a CPU resource allocation algorithm, dubbed \ac{PIRS},
based on \ac{ANBS} in Game Theory, which focuses on allocating idle resources in a proper proportion for each vehicle in every cooperation round. 
We have carried out simulations to investigate the performance of \ac{PIRS} and then compared the performance of \ac{PIRS} with two benchmark algorithms, \ac{AIRS} and \ac{NCS}.
The results of our simulations show that our proposed approach performs better in all aspects considered: it provides a lower amount of failed tasks, a lower amount of offloading to the edge and remote cloud, and higher success in executed task lengths than \ac{AIRS} and \ac{NCS}, especially when the density of the vehicles is lower.

\begin{acronym} 
\acro{VM}{virtual machine}
\acro{3GPP}{Third Generation Partnership Project}
\acro{ASIC}{Application-specific Integrated Circuit}
\acro{ANN}{Artificial Neural Network}
\acro{AP}{Access Point}
\acro{BBU}{Baseband Unit}
\acro{C-RAN}{Cloud Radio Access Network}
\acro{CPU}{Central Processing Unit}
\acro{C-RANs}{Cloud Radio Access Networks}
\acro{D-RAN}{Distributed Radio Access Network}
\acro{DNN}{Deep Neural Network}
\acro{FDD}{Frequency Division Duplex}
\acro{FTTx}{Fibre To The x}
\acro{GPP}{General Purpose Processor}
\acro{HD}{High Definition}
\acro{IoT}{Internet of Things}
\acro{LTE}{Long Term Evolution}
\acro{LXC}{Linux Container}
\acro{MCS}{Modulation and Coding Scheme}
\acro{MDP}{Markov Decision Process}
\acro{NFV}{Network Function Vitualisation}
\acro{OFDM}{Orthogonal Frequency Division Multiplexing}
\acro{OLT}{Optical Line Terminal}
\acro{ONU}{Optical Network Unit}
\acro{OTT}{Over-the-top}
\acro{PHY}{Physical}
\acro{PON}{Passive Optical Networks}
\acro{PRB}{Physical Resource Block}
\acro{QAM}{Quadrature Amplitude Modulation}
\acro{RAS}{Random-Action-Selection}
\acro{RRH}{Remote Radio Head}
\acro{SDN}{Software-Defined Networking}
\acro{SDR}{Software-Defined Radio}
\acro{TBS}{Transport Block Size}
\acro{TDM}{Time Division Multiplexing}
\acro{vBBU}{virtual Baseband Unit}
\acro{VNO}{Virtual Network Operator}
\acro{vOLT}{vitual Optical Line Terminal}
\acro{WAN}{Wide Area Network}
\acro{WDM}{Wavelength Division Multiplexing}
\acro{UE}{User Equipment}

\acro{3GPP}{Third Generation Partnership Project}
\acro{ANBS}{Asymmetric Nash Bargaining Solution}
\acro{ASI}{Artificial Swarm Intelligence}
\acro{ASIC}{Application-specific Integrated Circuit}
\acro{ANN}{Artificial Neural Network}
\acro{BBU}{Baseband Unit}

\acro{DNN}{Deep Neural Network}
\acro{EMP}{Evolutionary Multi-robots Personality}
\acro{EMP-A}{EMP with Asymmetric Nash Bargaining strategy}
\acro{EMP-F}{EMP with Fixed step strategy}
\acro{FDD}{Frequency Division Duplex}
\acro{FTTx}{Fibre To The x}
\acro{GPP}{General Purpose Processor}
\acro{HD}{High Definition}
\acro{IaaS}{Infrastructure as a Service}
\acro{IoT}{Internet of Things}
\acro{LTE}{Long Term Evolution}
\acro{LXC}{Linux Container}
\acro{NBS}{Nash Bargaining Solution}
\acro{NBSS}{Nash Bargaining Solution Sharing}
\acro{PT}{Personality Traits}
\acro{PA}{Probability of selecting Actions}
\acro{QoE}{Quality of Experience}
\acro{SEMP}{Single Evolution Multi-robots Personality}
\acro{SDN}{Software-Defined Networking}
\acro{SDR}{Software-Defined Radio}
\acro{TBS}{Transport Block Size}
\acro{TDM}{Time Division Multiplexing}
\acro{vBBU}{virtual Baseband Unit}
\acro{VNO}{Virtual Network Operator}
\acro{vOLT}{vitual Optical Line Terminal}
\acro{WDM}{Wavelength Division Multiplexing}
\acro{SUMO}{Simulation of Urban MObility}

\acro{NN}{Neural Network}
\acro{MEC}{Mobile Edge Cloud}
\acro{RAN}{Radio Access Network}
\acro{V2VEMP}{V2V EMP share strategy}
\acro{V2VSA}{V2V all spare value share strategy}
\acro{DSRC}{Dedicated Short Range Communications}
\acro{V2V}{Vehicle to Vehicle}
\acro{V2I}{Vehicle to Infrastructure}
\acro{VANET}{Vehicular Ad Hoc Network}
\acro{ISA}{Intelligent Speed Advisory}
\acro{IPA}{Intelligent Parking Advisory}
\acro{ADAS}{Advanced Driver Assistant System}
\acro{DSAS}{Distributed Speed Advisory System}
\acro{ICEVs}{Internal Combustion Engine Vehicles}
\acro{AI}{Artificial Intelligence}
\acro{ES}{Edge Server}
\acro{EN}{Edge Node}
\acro{OBU}{On-Board Units}
\acro{CO}{Central Office}
\acro{CSC}{Central SDN Controller}
\acro{GIPS}{giga instructions per second}
\acro{RSU}{Road Side Units}

\acro{MNO}{Mobile Network Operator}
\acro{ETSI}{European Telecommunications Standards Institute}
\acro{eMBB}{(enhanced Mobile Broad Band}
\acro{URLLC}{Ultra Reliability and Low latency Communications}
\acro{mMTC}{massive Machine Type Communications}
\acro{IaaS}{Infrastructure-as-a-Service}
\acro{NFV}{Network Function Virtualization}
\acro{NFVI}{Network Function Virtualization Infrastructure}
\acro{MEC-MANO}{MEC management and orchestration}
\acro{NFV-MANO}{NFV management and orchestration}
\acro{MANO}{management and orchestration}
\acro{CAPEX}{Capital Expenditure}
\acro{OPEX}{Operational Expenditure}
\acro{VM}{Virtual Machine}
\acro{SP}{service provider}
\acro{AIRS}{All Idle Resource Strategy}
\acro{PIRS}{Partial Idle Resource Strategy}
\acro{NCS}{Non-Cooperation Strategy }

\acro{EC}{Edge Computing}
\acro{VM}{Virtual Machine}
\acro{IoT}{Internet of Things}
\acro{UE}{User Equipment}

\end{acronym}

\section*{Acknowledgement}
\small
Financial support from Science Foundation Ireland (SFI) grants 17/CDA/4760, 18/RI/5721 and 13/RC/2077\_p2 is acknowledged.

\bibliographystyle{IEEEtran}
\bibliography{bibliography}
\end{document}